# Dirac-Surface-State-Dominated Spin to Charge Current Conversion in the Topological Insulator $(Bi_{0.22}Sb_{0.78})_2Te_3$ Films at Room Temperature

J. B. S. Mendes[1*], O. Alves-Santos[2], J. Holanda[2], R. P. Loreto[1], C. I. L. de Araujo[1], Cui-Zu Chang[3,4], J. S. Moodera[3,5], A. Azevedo[2], and S. M. Rezende[2]

[1]Departamento de Física, Universidade Federal de Viçosa, 36570-900, Viçosa, MG, Brazil

[2]Departamento de Física, Universidade Federal de Pernambuco, 50670-901, Recife, PE, Brazil

[3]Francis Bitter Magnet Lab, Massachusetts Institute of Technology, Cambridge, MA 02139, USA

[4]Department of Physics, The Pennsylvania State University, University Park, PA16802, USA

[5]Department of Physics, Massachusetts Institute of Technology, Cambridge, MA 02139, USA

We report the spin to charge current conversation in an intrinsic topological insulator (TI) $(Bi_{0.22}Sb_{0.78})_2Te_3$ film at room temperature. The spin currents are generated in a thin layer of permalloy (Py) by two different processes, spin pumping (SPE) and spin Seebeck effects (SSE). In the first we use microwave-driven ferromagnetic resonance of the Py film to generate a SPE spin current that is injected into the TI $(Bi_{0.22}Sb_{0.78})_2Te_3$ layer in direct contact with Py. In the second we use the SSE in the longitudinal configuration in Py without contamination by the Nernst effect made possible with a thin NiO layer between the Py and $(Bi_{0.22}Sb_{0.78})_2Te_3$ layers. The spin-to-charge current conversion is attributed to the inverse Edelstein effect (IEE) made possible by the spin-momentum locking in the electron Fermi contours due to the Rashba field. The measurements by the two techniques yield very similar values for the IEE parameter, which are larger than the reported values in the previous studies on topological insulators.

Topological Insulators (TIs) constitute a novel state of matter, which have been the subject of intensive investigations in condensed matter physics in the last decade. They are a new class of quantum materials that present insulating bulk, but metallic dissipationless surface states topologically protected by time reversal symmetry, opening several possibilities for practical applications in many scientific arenas including spintronics, quantum computation, magnetic monopoles and highly correlated electron systems [1-4]. More recently, it has been shown that TI-particles behave as optically induced oscillators in an optical tweezers [5]. The surface states are characterized by a single gapless Dirac cone and exhibit the remarkable spin-momentum locking: charge carrier move in such a way that their momenta are always perpendicular to their spin [4,6]. In addition, topological insulators have strong spin-orbit coupling (SOC) and as well have large spin-torque which are essential for efficient spin-charge conversion [7-9].

In turn, the conversion of charge currents into spin currents, and vice versa, are key phenomena for encoding and decoding information carried by electron spins in the active field of spintronics. Until recently, the only known mechanisms for conversion in both directions were the spin Hall effect (SHE) and its Onsager reciprocal, the inverse spin Hall effect (ISHE), that rely on electron scattering processes with spin-orbit interaction in 3D materials [10-14]. Studies of the spin-to-charge conversion by the ISHE have been conducted in metallic films with heavy elements, such as paramagnetic Pt, Pd, and Ta [15-27], ferromagnetic Py [28], antiferromagnetic materials such as IrMn and PtMn [29,30], and semiconductors [31-37]. Recent developments in thin-film growth techniques have made possible the fabrication of samples with atomically flat surfaces and interfaces that have led to the observation of new phenomena induced by SOC in 2D systems [38-40]. Among them are the Edelstein effect, predicted some time ago [41], and its Onsager reciprocal, the inverse Edelstein effect, that enable new means to convert charge currents into spin currents, and vice versa.

The direct Edelstein effect and the inverse Edelstein effect (IEE) are made possible by the Rashba effect that arises from SOC and broken inversion symmetry at material surfaces and interfaces [42-45]. The Rashba field produces spin-momentum locking in the electron Fermi contours that enables the conversion between spin and charge currents. The conversion of spin currents produced by ferromagnetic resonance (FMR) spin pumping into charge currents due to the IEE has been observed at Bi/Ag interfaces [46], single-layer graphene [47] and in a few TIs [8,48-50]. In this *Rapid Communication*, we report the observation of spin-to-charge current conversion by means of the IEE in the topological insulator $(Bi_{0.22}Sb_{0.78})_2Te_3$ at room temperature. The spin currents are generated by two different arrangements, microwave-driven spin pumping and the spin Seebeck effect.

The experiments were carried out with two sample structures: i) consisting of the TI grown on a 0.5 mm thick sapphire (0001) substrate and with a $Ni_{81}Fe_{19}$ (permalloy-Py) top layer, ii) a trilayer in which a NiO layer is grown between the TI and the Py layers. In both, we have used a commercial 0.5 mm thick (0001) sapphire substrate onto which the TI is grown as follows. After high temperature annealing (~800 ºC) of the sapphire substrate, a six-quintuple layer (QL)-thick $(Bi_xSb_{1-x})_2Te_3$ film is grown on top at a temperature ~230 ºC in a custom-built ultrahigh vacuum molecular beam epitaxy (MBE) system and capped by a 3-nm-thick epitaxial Te layer. X-ray diffraction patterns confirm the high crystalline, single phase quality of the films, with growth along the *c* axis (see in the Supplemental Material more details about the conditions of growth and the crystallographic structure of the (Bi,Sb)Te films [51]). We have chosen the Bi concentration $x=0.22$ to locate the Fermi level close to the Dirac point [52-54]. The Py layer is deposited by DC magnetron sputtering, either directly on the TI film or separated by an insulating NiO layer, grown by RF sputtering at 160 ºC. The Py and NiO films were deposited in a 3 mTorr argon atmosphere in the sputter-up configuration, with the substrate at a distance of 9 cm from the target, and with a deposition rate fixed in 1 Å/s and 0.3 Å/s, respectively. Therefore, the Py and NiO layers were gently deposited over the TI to minimize any detrimental effect on the surface chemistry. Finally, two silver electrodes were attached to the ends of the TI layer for measuring the induced voltages.

For the ferromagnetic resonance and spin pumping experiments the sample was mounted on the tip of a PVC rod and inserted through a hole drilled in the center of the back wall of a rectangular microwave cavity operating in the $TE_{102}$ mode, at a frequency of 9.4 GHz with a *Q* factor of 2000. The sample is slightly inserted into the cavity in the plane of the back wall, in a position of maximum *rf* magnetic field and minimum *rf* electric field to avoid the generation of galvanic effects driven by the electric field. With this arrangement the static magnetic field *H* and the microwave field $h_{rf}$ are in the film plane and kept perpendicular to each other as the sample is rotated for the measurements of the angular dependence of the FMR spectra and the *dc* voltage induced by the magnetization precession. Field scan spectra of the derivative of the microwave absorption d*P*/d*H* are obtained by modulating the field at 1.2 kHz and using lock-in detection. All FMR and voltage measurements were taken at room temperature.

Figure 1(a) shows a schematic illustration of the 6QL $(Bi_{0.22}Sb_{0.78})_2Te_3$/Py (12 nm) bilayer sample used in the SPE experiments, that has length 3 mm and width 1.5 mm. The Py films have in-plane magnetization and thus the magnetic proximity effect is expected to shift the Dirac cone sideways along the momentum direction and does not open an exchange gap (i.e. in our heterostructures, the Dirac cone of the TI film will be preserved). Figure 1(b) shows the FMR absorption spectrum of the Py layer in contact with the TI film measured with microwave power of 24 mW. The FMR line has the shape of a Lorentzian derivative with peak-to-peak linewidth of 38.1 Oe, corresponding to a half-width at half-maximum (HWHM) linewidth of $\Delta H = 33.0$ Oe. As shown in the inset of Fig. 1(b), an identical Py layer deposited on a Si substrate has linewidth $\Delta H_{Py} = 28.0$ Oe, showing that the contact of the TI layer produces an additional damping due to the spin pumping process [55,56], similar that observed in Pt/Py bilayers [17,19]. Figure 1(c) shows the field ($H$) scan $dc$ voltage measured directly with a nanovoltmeter connected by copper wires to the electrodes, for a microwave power of 24 mW, for three angles of the in-plane field. For $\phi = 0°$ the voltage lineshape is the superposition of symmetric and antisymmetric components, changes sign with inversion of the field, and vanishes for the field along the sample strip $\phi = 90°$.

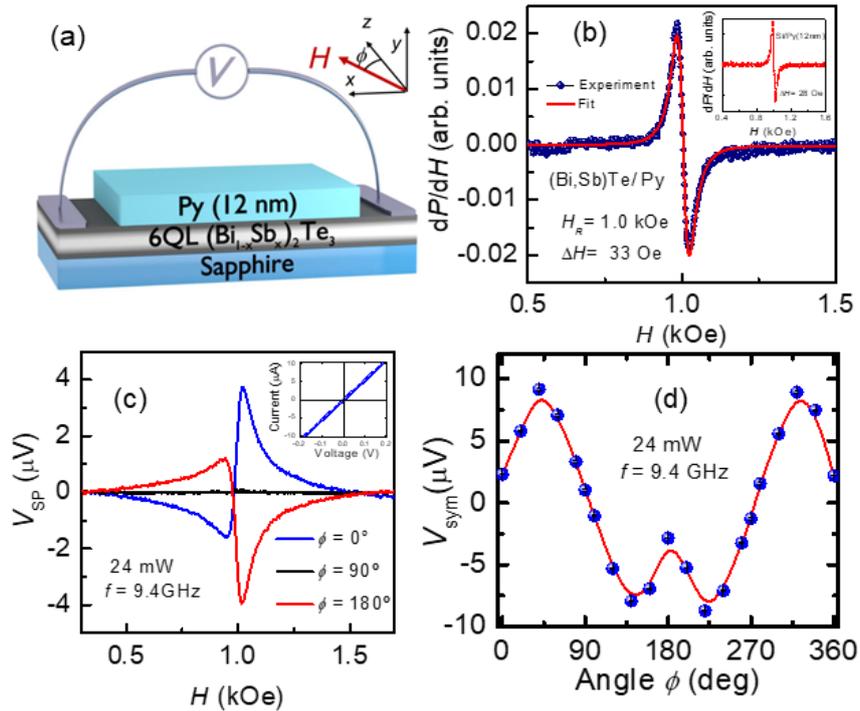

**Figure 1**. (color online) (a) Sketch of the bilayer sample 6QL (Bi$_{0.22}$Sb$_{0.78}$)$_2$Te$_3$/Py(12 nm) and coordinate system, where $\phi$ is the in-plane angle. (b) FMR absorption derivative versus magnetic field $H$ measured at 9.4 GHz and microwave power of 24 mW. Inset shows the FMR spectrum for a single Py (12 nm) layer on a Si substrate. (c) Voltage measured between the electrodes for three angles of the in-plane field, with the same microwave frequency and power as in (b). Inset shows the $I$-$V$ curve of the (Bi$_{0.22}$Sb$_{0.78}$)$_2$Te$_3$/Py structure demonstrating the formation of Ohmic contacts between the electrodes. (d) Angular dependence of the symmetric (peak) component of the voltage line. The experimental data are represented by the solid circle symbols and the theoretical fit by the solid curve.

The field dependence voltage $V(H)$ measured between the electrodes can be described by the sum of two components, $V(H) = V_{sym} L(H - H_R) + V_{asym} D(H - H_R)$, where $L(H - H_R)$ is the (symmetric) Lorentzian function and $D(H - H_R)$ is the (antisymmetric) Lorentzian derivative centered about the FMR resonance field $H_R$. The voltage lineshape measured as function of the field angle $\phi$ can be fit with the expression

$$V(H,\phi) = V_Q^{peak} L(H - H_R) \cos\phi + [V_{CL}^{sym} L(H - H_R) + V_{CL}^{asym} D(H - H_R)] \sin 2\phi \sin\phi, \quad (1)$$

where $V_{CL}^{sym}$ and $V_{CL}^{asym}$ denote the amplitudes of the symmetric and antisymmetric components of the classical contributions, such as the galvanic effect, or spin rectification, generated in the Py layer [19], and $V_Q^{peak}$ is the peak value of the symmetric contribution to the voltage of quantum origin which will be discussed later. Figure 1 (d) shows the measured angle dependence of the symmetric component, which is the one of interest here, and a solid curve representing the fit obtained with Eq. (1).

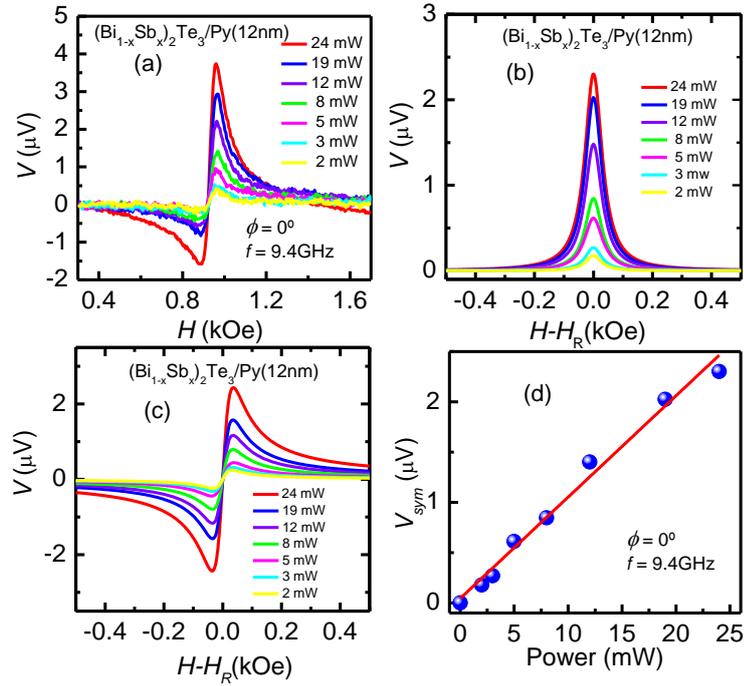

**Figure 2**. (color online) (a) Voltage measured between the electrodes at the field angle $\phi = 0°$ for several microwave power levels as indicated. (b) and (c) Variation with power of the symmetric and antisymmetric components of the voltage obtained by the fitting of Lorentzian a Lorentzian derivative functions to the lineshapes in (a). (d) Power dependence of the measured symmetric peak component of the voltage at $\phi = 0°$.

Figure 2(a) shows the voltage lineshapes measured at several power levels and Figs. 2(b) and 2(c) show the corresponding symmetric and antisymmetric components of the lineshapes, obtained by fitting the sum of a Lorentzian function and a Lorentzian derivative to the data. Figure 2 (d) shows that the symmetric component at $\phi = 0°$, which is $V_Q^{peak}$, exhibits a linear dependence with the microwave power. From Figs. 1(d) and 2(b) we have the value for the voltage of quantum origin at $\phi = 0°$, $V_Q^{peak} = 2.3 \mu V$, for a microwave power of 24 mW, part of which is due to the magnonic charge pumping (MCP) that is produced in a single Py layer [57]. In order to separate the contributions, we have measured the voltage induced in a single layer of Py(12 nm) on Si substrate in the same conditions of the data in Figs. (1) and (2). For a microwave power of 24 mW the symmetric component at $\phi = 0$ has a peak value of 0.4 µV.

The most important source for the symmetric component of the voltage, and the one of interest here, is the conversion of the spin current produced by spin pumping into charge current in the TI layer. As is well known [12-19], in a ferromagnetic (FM) layer under ferromagnetic resonance, the precessing magnetization generates a spin current density (in units of charge/time area) at the FM/TI interface given by

$$J_S = \frac{e \omega p g_{eff}^{\uparrow\downarrow}}{8\pi} \left( \frac{h_{rf}}{\Delta H} \right)^2 L(H - H_R), \tag{2}$$

where $e$ is the electron charge, $g_{eff}^{\uparrow\downarrow}$ is the real part of the effective spin mixing conductance at the interface that takes into account the spin-pumped and back-flow spin currents [12-19,55,56], $\omega$ and $h_{rf}$ are, respectively, the frequency and amplitude of the driving microwave magnetic field, and $p$ is the precession ellipticity factor $p = 4(\omega/\gamma)(H_R + 4\pi M_{eff})/(2H_R + 4\pi M_{eff})^2$, where $4\pi M_{eff}$ is the effective magnetization that appears in the expression for the FMR frequency $f = \gamma [H_R(H_R + 4\pi M_{eff})]^{1/2}$. The spin current produced by the FMR spin pumping flows through the FM/TI interface into the TI layer. We follow Refs. [8, 44-50] and interpret the spin-to-charge conversion in the TI layer as arising from the inverse Edelstein effect (IEE), that has its origin in the spin-momentum locking in the Fermi contours due to the Rashba SOC interaction. The 3D spin current in Eq. (2) flows into the TI layer and is converted by the IEE into a lateral charge current with a 2D density $j_C = (2e/\hbar) \lambda_{IEE} J_S$, where $\lambda_{IEE}$ is a coefficient characterizing the IEE, with dimension of length and proportional to the Rashba coefficient, and hence to the magnitude of the SOC [8,44-46]. The measured voltage is related to this current density by $V_{IEE} = R_S w j_C$, where $R_S$ is the shunt resistance, $w$ the width of the (Bi,Sb)Te/Py bilayer in the length of the Py layer and $j_C$ has units of A/m.

In order to obtain the IEE length from the experimental data, we need initially to calculate the SPE spin current. The real part of the spin mixing conductance of the (Bi,Sb)Te/Py interface that enters in Eq. (2) can be inferred from the broadening of the FMR linewidth due to the spin pumping process using $g_{eff}^{\uparrow\downarrow} = (4\pi M_0 t_{Py}/\hbar\omega)(\Delta H - \Delta H_{Py})$ [12-14,17,55,56]. With $4\pi M_0 = 11$ kG, $t_{Py} = 12$ nm, $\omega/2\pi = 9.4$ GHz, we find that the additional linewidth of 5 Oe measured in Py due to the contact with the TI layer corresponds to $g_{eff}^{\uparrow\downarrow} = 1.0 \times 10^{19}$ m$^{-2}$, a value similar to the one for Py/Pt interfaces [12-19]. The amplitude of the microwave field in Eq. (1), in Oersted, is related to the incident power $P_i$, in watt, by $h_{rf} = 1.776(P_i)^{1/2}$, calculated for a microwave cavity made with a shorted standard X-band rectangular waveguide, operating in the $TE_{102}$ mode with $Q$ factor of 2000, at a frequency of 9.4 GHz. Using these values, we obtain for $P_i = 24$ mW, $H = H_R$, the spin current density at the interface produced by the FMR spin pumping $J_S = 2.3 \times 10^5$ A/m$^2$. The charge current density due to the conversion from the spin current by the IEE, given by $j_C = V_{IEE}^{peak}/(wR_S)$, corresponding to the measured voltage of $V_{IEE}^{peak} = 1.9 \,\mu$V, considering that the shunt resistance is approximately the one of the Py layer, $R_S = 71\,\Omega$, and $w = 1.5$ mm, is $j_C = 1.7 \times 10^5$ A/m. Thus, the IEE length $\lambda_{IEE} = j_C/J_S$ obtained from the spin pumping measurements is $\lambda_{IEE} = 0.075$ nm.

The spin pumping origin of the spin current pumped into the TI was confirmed by using a broadband microwave microstrip setup and measuring the voltage between the electrodes on the TI strip with scanning $H$ for several frequencies, keeping constant the incident power at the $P_i = 28$ mW. With this power level, the *rf* magnetic field produced in the sample placed on a copper microstrip 0.5 mm wide with characteristic impedance $Z_0 = 50\,\Omega$, is very similar to the one in the microwave cavity with $Q$=2000 with power 24 mW previously described. The field dependencies of the voltages measured at several frequencies are shown in the lower panel of Fig. 3 with Lorentzian fits. One clearly see the broadening of the voltage lines with increasing frequency, which is a characteristic feature of the spin pumping damping [55,56]. The upper panel in Fig. 3 shows the variation of the measured field value for the peak voltage, that is the FMR field $H_R$, with the driving frequency. In order to compare the voltages measured at the various frequencies we need to consider that the peak value of the pumped spin current varies inversely with the FMR linewidth squared, $\Delta H^2$, as in Eq. (2). Thus we introduce the normalized peak voltage, defined by $V^* = V(\Delta H / \Delta H_5)^2$, where $V$ and $\Delta H$ are, respectively, the peak voltage and the linewidth measured at a frequency $f$, and $\Delta H_5$ is the linewidth at 5 GHz. The linear increase of the normalized peak voltage shown in the inset of the upper panel of Fig. 3 provides another evidence of the spin pumping origin of voltage induced in the TI layer.

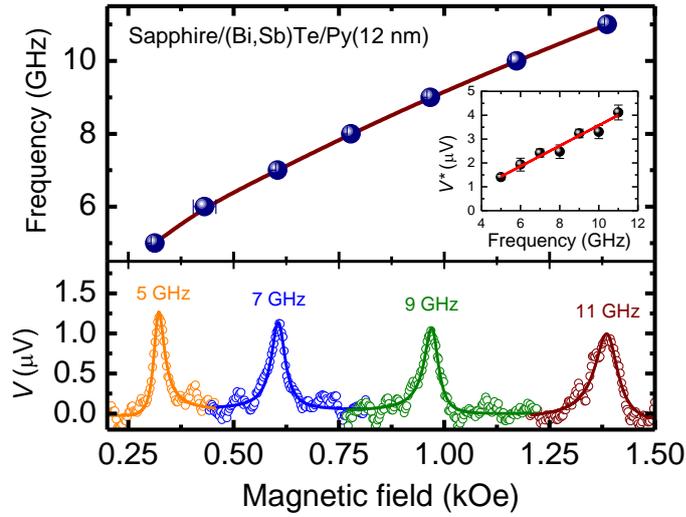

**Figure 3**. (color online) Lower panel: the voltage measured between the electrodes on the TI strip at the field angle $\phi = 0°$ for several microwave driving frequencies as indicated, and power 28 mW. Upper panel: Driving frequency versus the field value for the peak voltage. Inset shows the variation with frequency of the peak voltage normalized by the FMR linewidth squared referred to the value at 5 GHz.

In order to confirm the spin-to-charge current conversion in $(Bi_{1-x}Sb_x)_2Te_3$ by the IEE, we have used another process to generate spin currents, namely, the spin Seebeck effect (SSE). We used a sample arrangement illustrated Fig. 4(a), in which a 5 nm thick NiO layer provides electrical isolation between the TI and Py films. The Py layer has width 1.0 mm, smaller than the NiO and TI layers, to avoid possible contacts at the edges. A commercial Peltier module, of width 4 mm, is used to heat or cool the side of the Py layer while the substrate is maintained in thermal contact with a copper block at room temperature. The temperature difference $\Delta T$ across the sample is calibrated as a function of the current in the Peltier module by means of a differential thermocouple. The temperature gradient perpendicular to the Py layer has two effects: One is to generate a voltage along the layer by means of the classical anomalous Nernst effect (ANE) [58-61]; the other is to generate a spin current across the Py layer by the longitudinal spin Seebeck effect [61-66]. Since NiO is a room temperature antiferromagnet, it blocks the flow of charge current but transports spin currents [66-69], thus allowing the measurements of the voltage generated in the TI layer separated from the voltage induced in the Py layer by the ANE. Figure 4(b) shows the voltages measured between the two electrodes in the $(Bi_{1-x}Sb_x)_2Te_3$ layer that are produced by the electric current resulting from the IEE spin-to-charge conversion of the spin current generated by the spin Seebeck effect in the Py layer that is injected into the Py/NiO interface and transported by the magnons in the NiO layer. The magnetic field dependencies of the SSE-IEE voltages in the TI layer for several values of the temperature difference $\Delta T$ across the sample structure are shown in Fig. 4 (b). Note that $\Delta T > 0$ corresponds to the Peltier module warmer than the substrate. The data have the shape of the hysteresis curve of Py with very small coercitivity in the field scale of the measurements. The change in the voltage sign with the field reversal is due to the sign change of the spin polarization. Figure 4 (c) shows the measured variation of the voltage plateau with the temperature difference $\Delta T$ for applied fields of $H = \pm 0.4$ kOe. The linear dependence of $V_{SSE}$ on $\Delta T$ results from the fact that the spin current generated by the LSSE in Py is proportional to the temperature gradient across the Py layer [66]. In order to calculate the IEE parameter obtained from the SSE-IEE measurements we use data for the SSE in Si/Py/NiO/Pt in Ref. [66] and rescale the numbers accordingly. The value of the charge current measured in the Pt layer in the sample Si/Py(20 nm)/NiO(5 nm)/Pt (6 nm) for $\Delta T = 12$ K is 24.3 nA, corresponding to a current density in the Pt layer of width 2.5 mm and thickness 6 nm of $J_C = 1.6 \times 10^3$ A/m$^2$ [66]. Considering for the spin Hall angle of Pt the value $\theta_{SH} = 0.05$ [12-14], this gives for the spin current density reaching the Pt after attenuation in the NiO layer the value $J_S = 3.2 \times 10^4$ A/m$^2$ for $\Delta T = 12$ K. However, we must consider that the temperature gradients in the Py layer in the Si/Py/NiO/Pt and in the Al$_2$O$_3$/TI/NiO/Py samples are different. The temperature gradient across the Py layer in the Si/Py/NiO/Pt sample is given by $\nabla T_{Py} \approx (K_{Si}/K_{Py})(\Delta T/t_S)$, where $K_{Py}$ and $K_{Si}$ are the thermal conductivities of Py and Si, and $t_S$ is the sample thickness. For $\Delta T = 12$ K, $t_S = 0.4$ mm, $K_{Si} = 148$ W/(K m), $K_{Py} = 46.4$ W/(K m), $\nabla T_{Py} \approx 957$ K/cm. On the other hand, in the Al$_2$O$_3$/TI/NiO/Py sample, the gradient corresponding to a temperature difference $\Delta T$ is $\nabla T_{Py} \approx (K_{Al2O3}/K_{Py})(\Delta T/t_S)$, which gives for $\Delta T = 12$ K, $t_S = 0.5$ mm, $K_{Al2O3} = 41.9$ W/(K m), $\nabla T_{Py} \approx 217$ K/cm. Thus, the same $\Delta T = 12$ K would generate in the TI layer a spin current density $J_S = 7.2 \times 10^3$ A/m$^2$. From the data in Fig. 4 (c), for $\Delta T = 12$ K the voltage in the TI layer is $V_{SSE} = 5.9 \mu$V, which, for a resistance of $R = 1.08 \times 10^4$ $\Omega$ of the TI in the length of the Peltier module in a width of 1 mm corresponds to a 2D current density of $j_C = 5.46 \times 10^{-7}$ A/m. In turn, the resistance between the Py film and the TI layer is $R \geq 1.0 \times 10^6$ $\Omega$. Therefore, the resistance measurements give us assurance that the NiO layer (5nm) is a good electrical insulator. Thus, the IEE parameter $\lambda_{IEE} = j_C/J_S$ obtained from the SSE measurements is $\lambda_{IEE} = 0.076$ nm, which is nearly the same obtained from the SPE measurements.

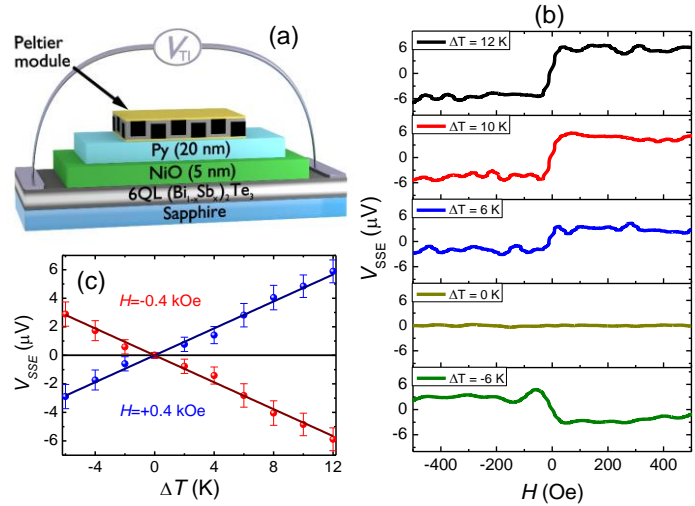

**Figure 4**. (color online) (a) Schematic illustration of the sample 6QL $(Bi_{0.22}Sb_{0.78})_2Te_3$/NiO/Py(12 nm) structure used to measure the voltages generated by the ANE and SSE. (b) Variation with magnetic field of the SSE-IEE voltage measured in the TI layer, created by the combined SSE in the Py layer and IEE in the TI, for several values of $\Delta T$ as indicated. Positive $\Delta T$ corresponds to the Peltier module warmer than the substrate. (c) Variation with temperature difference of the SSE-IEE voltage measured with $H=0.4$ kOe, in two field directions.

In summary, we have demonstrated the conversion of a spin current into a charge current in the topological insulator (Bi$_{0.22}$Sb$_{0.78}$)$_2$Te$_3$ at room temperature, that is attributed to the inverse Edelstein effect (IEE) made possible by the spin-momentum locking in the electron Fermi contours due to the Rashba field. The spin currents were generated in a thin layer of permalloy by two different processes, spin pumping (SPE) and spin Seebeck effects (SSE). In the former we have used microwave-driven ferromagnetic resonance of the Py film to generate a spin current that is injected into the TI film in direct contact with Py. In the latter we have used the SSE in the longitudinal configuration in Py with no contamination by the Nernst effect made possible with the use of a thin NiO layer between the Py and TI layers. The results of the two measurements yield nearly identical values for the IEE coefficient, 0.076 nm, which is about twice the value measured for the topological insulator $Bi_2Se_3$ (6 QL) and about four times larger than the value for $(Bi,Sb)_2Te_3$ (6 QL) [50].

## ACKNOWLEDGEMENTS

This research was supported in Brazil by Conselho Nacional de Desenvolvimento Científico e Tecnológico (CNPq), Coordenação de Aperfeiçoamento de Pessoal de Nível Superior (CAPES), Financiadora de Estudos e Projetos (FINEP), Fundação de Amparo à Pesquisa do Estado de Minas Gerais (FAPEMIG), and Fundação de Amparo à Ciência e Tecnologia do Estado de Pernambuco (FACEPE). CZC and JSM acknowledge the support from NSF Grants No. DMR-1207469, DMR-1700137, No. DMR- 0819762 (MIT MRSEC), ONR Grant No. N00014-16-1-2657, and the STC Center for Integrated Quantum Materials under NSF Grant No. DMR-1231319. CZC thanks the support from the startup provided by Penn State.

# Supplementary information:
# Dirac-Surface-State-Dominated Spin to Charge Current Conversion in the Topological Insulator $(Bi_{0.22}Sb_{0.78})_2Te_3$ Films at Room Temperature

J. B. S. Mendes[1*], O. Alves-Santos[2], J. Holanda[2], R. P. Loreto[1], C. I. L. de Araujo[1], Cui-Zu. Chang[3,4], J. S. Moodera[3,5], A. Azevedo[2], and S. M. Rezende[2]

[1]Departamento de Física, Universidade Federal de Viçosa, 36570-900, Viçosa, MG, Brazil
[2]Departamento de Física, Universidade Federal de Pernambuco, 50670-901, Recife, PE, Brazil
[3]Francis Bitter Magnet Lab, MIT, Cambridge, Massachusetts 02139, USA
[4]Department of Physics, The Pennsylvania State University, University Park, PA16802, USA
[5]Department of Physics, MIT, Cambridge, Massachusetts 02139, USA

## I. The growth of topological insulator $(Bi_{1-x}Sb_x)_2Te_3$ thin films

The topological insulator (TI) thin film growth was performed using a custom-built ultrahigh vacuum molecular beam epitaxy (MBE) system. The insulating heat-treated sapphire (0001) was used as substrate and has been outgassed at 800°C for one hour before the deposition. The Bi, Sb and Te were evaporated from Knudsen effusion cells using high purity Bi(99.999%), Sb(99.9999%) and Te (99.9999%). During the growth, the substrate was maintained at 230 °C. To reduce Te vacancies, the growth is kept in Te-rich condition (the flux ratio of Te per (Bi +Sb) was set to approximately ~10). The Bi, Sb and Te concentrations in the films were determined by their ratio, obtained in situ during growth using properly calibrated quartz crystal monitors. The growth rate for the TI films was ~0.2 quintuple layers (QLs) per minute. Following the growth, the TI films were annealed at 230°C for 30 minutes to improve the crystal quality before being cooled down to room temperature. The sharp and streaky 1×1 reflective high energy electron diffraction (RHEED) patterns (Figs. S1a and S1b) indicate the high single crystalline quality of 6QL $(Bi_{0.22}Sb_{0.78})_2Te_3$ film grown on sapphire (0001) substrate. To avoid possible contamination, a 3-nm thick epitaxial Te capping layer was deposited at room temperature on top of the TI films before taking the samples out of the MBE chamber for the device fabrications.

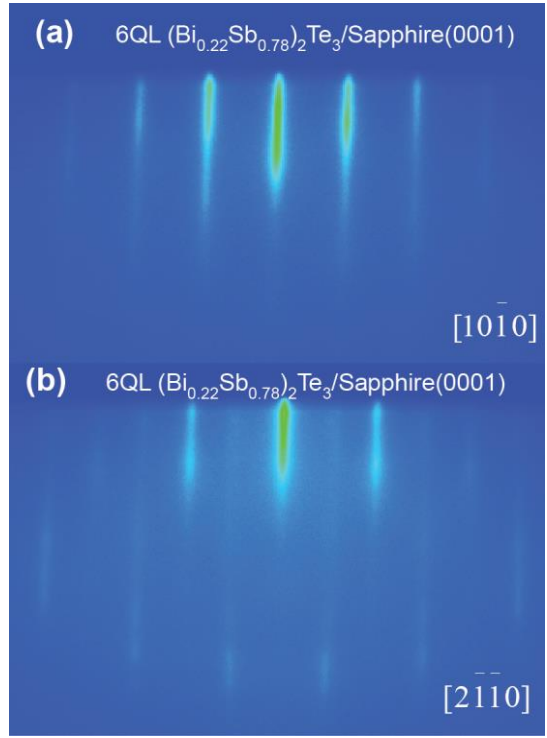

**Fig. S1.** (color online) RHEED patterns of 6 QL $(Bi_{0.22}Sb_{0.78})_2Te_3$ grown on sapphire (0001) with the electron beam parallel to along $[1\bar{1}\bar{1}0]$ (a) and $[2\bar{1}\bar{1}0]$ (b).

## II. X-ray diffraction of the TI films

The crystallographic structure of the $(Bi_{0.22}Sb_{0.78})_2Te_3$ film was assessed by X-ray diffraction (XRD) measurements. The XRD spectrum at high resolution detailing the position of the peaks of the TI film is shown in Fig. S2. The diffraction patterns were obtained at angles between 5° and 70° (2θ). The XRD pattern indicates that the reflections are only from (00$l$) family of planes of $(Bi_{1-x}Sb_x)_2Te_3$, with the sapphire (006) peak or with the (001) peak of the Te capping layer, suggesting that no impurity phase is present. The inset in Fig. S1 shows XRD rocking measured at 2$\theta$=17.24°, corresponding to the $(Bi_{0.22}Sb_{0.78})_2Te_3$ (006) peak. A Gaussian fit to the spectrum gives a full width at half maximum (FWHM) of only 0.07°, demonstrating that a perfectly aligned domain epitaxial film has grown. The XRD patterns were recorded using the Bruker D8 Discover Diffractometer equipped with the Cu Kα radiation (λ=1.5418Å).

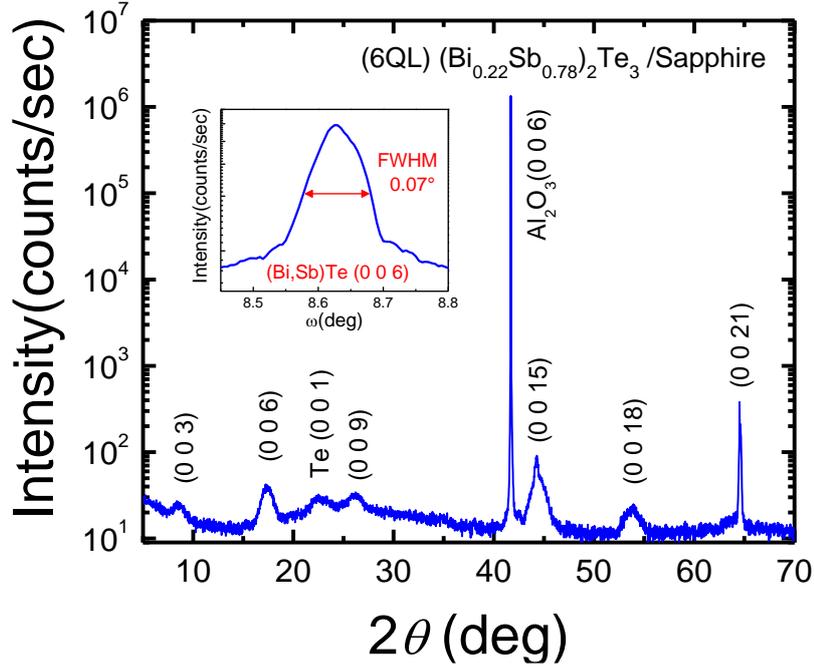

**Figure S2**. (color online) X-ray diffraction result of a typical 6 QL $(Bi_{0.22}Sb_{0.78})_2Te_3$ grown on sapphire $(Al_2O_3)$. The inset shows the XRD rocking curve, corresponding to the (Bi,Sb)Te (006) peak that gives a FWHM of 0.07°. The results of XRD measurements indicate that the present TI film is epitaxially grown on the sapphire substrate.

### III. Transport properties of TI films

Based on previous work we have chosen the Bi concentration x=0.22 to locate the Fermi level close to the Dirac point and so the resistivity in the bulk part should show insulating behavior. On this point, it is important to mention that the results involving the variation of Bi concentration to locate the Fermi level close to the Dirac point, have been explored in previous publications [52, 53 and 54]. In these previous publications, transport measurements (including resistivity, ordinary Hall measurements and anomalous Hall conductance) were also extensively explored. Figure S3 displays the temperature dependence of the longitudinal resistance, $R_{xx}$, for the TI-samples used in this manuscript. The resistance data indicate insulator-like behaviour for the samples. In the measure, $R_{xx}$ increases with decreasing temperature in almost the entire temperature range, reflecting the depletion of bulk carriers.

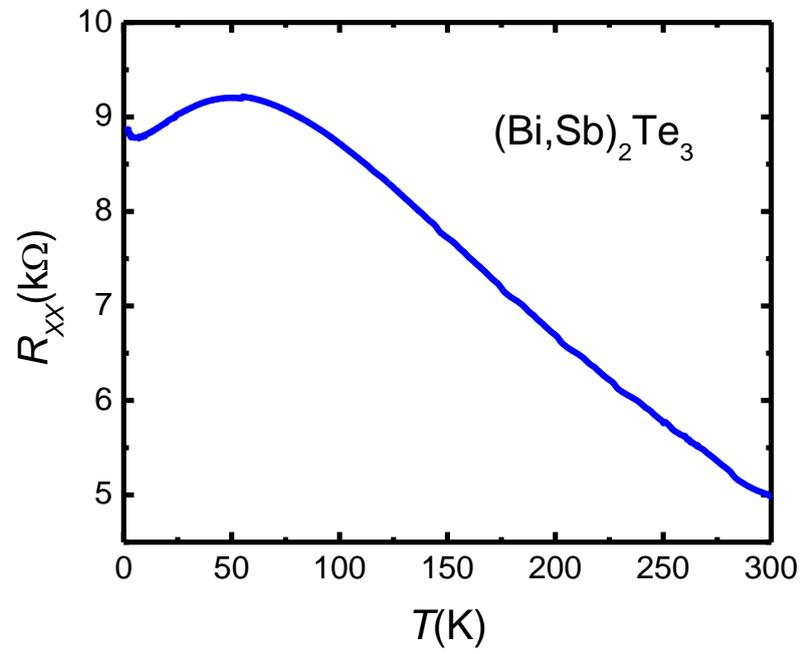

**Figure S3**. (color online) Temperature dependence of longitudinal resistance $R_{xx}$ 6 QL $(Bi_{0.22}Sb_{0.78})_2Te_3$ sample.